\documentclass[adp,fleqn]{w-art}
\usepackage{times}
\usepackage{w-thm}
\usepackage[]{graphicx}
\chardef\bslash=`\\ 

\begin{document}
\DOIsuffix{theDOIsuffix}
\Volume{12}
\Issue{1}
\Copyrightissue{01}
\Month{01}
\Year{2003}
\pagespan{1}{}
\Receiveddate{XXXX} \Reviseddate{XXXX} \Accepteddate{XXXX}
\Dateposted{XXXX}





\title[Non-standard LQC]{Non-standard loop quantum cosmology}


\author[W. \ Piechocki]{W{\l}odzimierz Piechocki\footnote{E-mail: {\sf piech@fuw.edu.pl}, Phone: (+48 22) 55 32 275,
     Fax: (+48 22) 62 16 085}}
\address{Department of Theoretical Physics, Institute for Nuclear
Studies,\\
 Hoza 69,  00-681 Warsaw, Poland }

\begin{abstract}
  We present results concerning the nature of the cosmological big bounce (BB) transition within the loop geometry underlying
  loop quantum cosmology (LQC). Our canonical quantization method
  is an alternative to the standard LQC.  An
  evolution parameter we use has clear interpretation both at
  classical and quantum levels.  The physical volume
  operator has discrete spectrum which is bounded from below.
  The minimum gap in the spectrum defines a quantum of the
  volume. The spectra of operators are parametrized by a free parameter to
  be determined.
\end{abstract}
\maketitle





\section{Introduction}
\label{sect1}

There are two  alternative methods of quantization of a
Hamiltonian system with  constraints: (i) Dirac's method - `first
quantize, then impose constraints', and (ii) non-Dirac's method -
`first solve constraints, then quantize'. We have two
corresponding methods in quantization of  cosmological models of
general relativity (GR) which make use of the so-called loop
geometry: (i) standard LQC - Dirac's method \cite{abl,boj}, and
(ii) non-standard LQC - method proposed recently \cite{ppw1,ppw2}.
The latter method corresponds to the reduced phase space
quantization of loop quantum gravity \cite{gies}.

In what follows we present quantization of  flat  FRW model with
massless scalar field by making use of the non-standard LQC.
Available results are the following: (1) within standard LQC -
classical Big-Bang is replaced by quantum Big-Bounce due  to
strong  quantum effects at the Planck scale
\cite{ash1,Dzierzak:2008dy,Bojowald:2008ik}; (2) within
non-standard LQC - modification of GR by loop geometry is
responsible for the resolution of the singularity, quantization
may lead to discrete spectra of physical observables
\cite{Malkiewicz:2009zd,Malkiewicz:2009xz}.

\section{Classical Level}
\label{sect2}

\subsection{Modified Hamiltonian}

The gravitational part of the Hamiltonian of the  flat FRW
universe with massless scalar field (in special gauges) is found
to be \cite{ppw1}
\begin{equation}\label{hamG}
H_g = - \gamma^{-2} \int_{\mathcal V} d^3 x ~N
e^{-1}\varepsilon_{ijk}
 E^{aj}E^{bk}  F^i_{ab}\, ,
\end{equation}
where  $\gamma$, Barbero-Immirzi parameter; $\mathcal V\subset
\Sigma$, elementary cell; $N$, lapse function;
$\varepsilon_{ijk}$, alternating tensor; $E^a_i $, density
weighted triad;  $ F^k_{ab} =
\partial_a A^k_b - \partial_b A^k_a + \epsilon^k_{ij} A^i_a
A^j_b$, curvature of $SU(2)$ connection $A^i_a$; $e:=\sqrt{|\det
E|}$;

Modification by loop geometry means approximation of $ F^k_{ab}$
as follows \cite{ppw1}
\begin{equation}\label{finite}
 F^k_{ab}(\lambda) \approx
-2\;Tr\;\Big(\frac{h^{(\lambda)}_{B_{ij}}-1}{\lambda^2
V_o^{2/3}}\Big)\;{\tau^k}\; ^o\omega^i_a  \;
^o\omega^j_a,\;\;\;\;\;\;\;F^k_{ab}= \lim_{\lambda\,\rightarrow
\,0}\, F^k_{ab}(\lambda),
\end{equation}
where the holonomy of the connection around the square  loop $
B_{ij}$, with sides length $\mu V_0^{1/3}$, reads
\begin{equation}\label{box}
   h^{(\mu)}_{B_{ij}} = h^{(\mu)}_i
h^{(\mu)}_j (h^{(\mu)}_i)^{-1}
(h^{(\mu)}_j)^{-1},\;\;\;\;\;h^{(\mu)}_k (c)  = \cos (\mu c/2) +
2\,\sin (\mu c/2)\;\tau_k,
\end{equation}
where $\tau_k = -i \sigma_k/2\;$ ($\sigma_k$ are the Pauli spin
matrices).

Making  use of   Thiemann's identity leads finally to
\begin{equation}\label{hamR}
    H_g = \lim_{\lambda\rightarrow \,0}\; H^{(\lambda)}_g ,
\end{equation}
where
\begin{equation}\label{hamL}
H^{(\lambda)}_g = - \frac{sgn(p)}{2\pi G \gamma^3 \lambda^3}
\sum_{ijk}\,N\, \varepsilon^{ijk}\, Tr \Big(h^{(\lambda)}_i
h^{(\lambda)}_j (h^{(\lambda)}_i)^{-1} (h^{(\lambda)}_j)^{-1}
h_k^{(\lambda)}\{(h_k^{(\lambda)})^{-1},V\}\Big),
\end{equation}
and where $V= |p|^{\frac{3}{2}}= a^3 V_0$ is the volume of the
elementary cell $\mathcal{V}$.  Variables $ c$ and $p$ determine
connections $A^k_a$ and  triads $E^a_k$: $A^k_a =
\,^o\omega^k_a\,c\,V_0^{-1/3} \,$ and $\,E^a_k =
\,^oe^a_k\,\sqrt{q_o}\,p\,V_0^{-2/3} $, where $\,c = \gamma
\,\dot{a}\,V_0^{1/3}$ and $\,|p| = a^2\,V_0^{2/3}$, $\{c,p\} = 8
\pi G \gamma /3$.

The total Hamiltonian for FRW universe with a massless scalar
field $\phi$ reads
\begin{equation}\label{ham}
   H = H_g + H_\phi,
\end{equation}
where $H_g$ is defined by (\ref{hamR}) and  $H_\phi = p^2_\phi
|p|^{-\frac{3}{2}}/2$, and where $\phi$ and $p_\phi$ are
elementary variables satisfying $\{\phi,p_\phi\} = 1$. The
relation $ H \approx 0$ defines the  physical phase space.

Making use of (\ref{box}) we calculate  (\ref{hamL}) and get the
modified  total Hamiltonian  corresponding to (\ref{ham})
\begin{equation}\label{regH}
   H^{(\lambda)}/N = -\frac{3}{8\pi G \gamma^2}\;\frac{\sin^2(\lambda
\beta)}{\lambda^2}\;v + \frac{p_{\phi}^2}{2\, v},\;\;\;\;\beta :=
\frac{c}{|p|^{1/2}},\;\;\;v := |p|^{3/2} .
\end{equation}

Equation (\ref{regH}) presents a  modified classical Hamiltonian.

\subsection{Observables}
A function, $\mathcal{O}: \mathcal{F}_{kin}^{(\lambda)}\rightarrow
R$, is a Dirac  observable  if
\begin{equation}\label{dirac}
\{\mathcal{O},H^{(\lambda)}\}= 0,\;\;\;\;\;\;\; \{\cdot,\cdot\}:=
4\pi G\gamma\;\bigg[ \frac{\partial \cdot}
    {\partial \beta} \frac{\partial \cdot}{\partial v} -
     \frac{\partial \cdot}{\partial v} \frac{\partial \cdot}{\partial \beta}\bigg] +
     \frac{\partial \cdot}{\partial \phi} \frac{\partial \cdot}{\partial p_\phi} -
     \frac{\partial \cdot}{\partial p_\phi} \frac{\partial \cdot}{\partial
     \phi}.
\end{equation}
Thus, $\mathcal{O}$ is solution to the equation
\begin{equation}\label{dir}
 \frac{\sin(\lambda\beta)}{\lambda}\,\frac{\partial
\mathcal{O}}{\partial\beta} - v \cos(\lambda\beta)\,\frac{\partial
\mathcal{O}}{\partial v} -
\frac{\kappa\gamma\,\textrm{sgn}(p_{\phi})}{4 \pi
G}\,\frac{\partial \mathcal{O}}{\partial\phi} = 0.
\end{equation}
Solutions to (\ref{dir}) are found to be \cite{ppw1}
\begin{equation}\label{obser1}
\mathcal{O}_1:= p_{\phi},\;\;\;\;\;\mathcal{O}_2:= \phi -
\frac{\textrm{sgn}(p_{\phi})}{3\kappa}\;\textrm{arth}\big(\cos(\lambda
\beta)\big),\;\;\;\;\; \mathcal{O}_3:= \textrm{sgn}(p_{\phi})\,v\,
\frac{\sin(\lambda \beta)}{\lambda}.
\end{equation}
Observables satisfy the Lie  algebra
\begin{equation}\label{ala1}
\{\mathcal{O}_2,\mathcal{O}_1\}=
1,\;\;\;\;\;\{\mathcal{O}_1,\mathcal{O}_3\}= 0,\;\;\;\;\;
\{\mathcal{O}_2,\mathcal{O}_3\}=  \gamma\kappa .
\end{equation}
Due to the constraint $H^{(\lambda)}=0$, we have $\mathcal{O}_3=
\gamma \kappa \,\mathcal{O}_1.$ Thus, in the physical phase space,
$\mathcal{F}_{phys}^{(\lambda)}$, we have only two observables
which satisfy the algebra
\begin{equation}\label{alg1}
\{\mathcal{O}_2,\mathcal{O}_1\}= 1.
\end{equation}

In what follows we consider functions which can be expressed in
terms of observables and an evolution parameter $ \phi$  so they
are not observables. They do become observables for each fixed
value of $ \phi$, since  in such case they are only functions of
observables:

\noindent The energy density of matter field is found to be
\cite{ppw1}
\begin{equation}\label{rho2}
\rho(\lambda,\phi) =
\frac{1}{2}\,\frac{1}{(\kappa\gamma\lambda)^2\,
    \cosh^2 3\kappa  (\phi- \mathcal{O}_2)}.
\end{equation}
The volume operator may be expressed as \cite{ppw1}
\begin{equation}\label{vol}
    v(\phi,\lambda) = \kappa\gamma\lambda\,
    |\mathcal{O}_1|\,\cosh3\kappa  (\phi-
    \mathcal{O}_2).
\end{equation}

\section{Quantum Level}

The energy density operator has been considered recently in
\cite{Malkiewicz:2009zd}. The spectrum of the quantum operator
corresponding to $\rho$ turns out to coincide with (\ref{rho2}).
The energy density operator is bounded and has continuous
spectrum.

In what follows we present quantization of the volume observable
\cite{Malkiewicz:2009xz}. The classical volume operator, $v$,
reads
\begin{equation}\label{vvol}
v = |w|,~~~w :=
\kappa\gamma\lambda\;\mathcal{O}_1\;\cosh3\kappa(\phi-
\mathcal{O}_2).
\end{equation}
Thus, quantization of $v$ reduces to the  quantization problem of
$w$:
\begin{equation}\label{c1}
\hat{w}\,f(x) :=
   \kappa\gamma\lambda\,\frac{1}{2}\,\big(
    \widehat{\mathcal{O}}_1\,\cosh3\kappa  (\phi-
    \widehat{\mathcal{O}}_2)
     + \cosh3\kappa  (\phi-
    \widehat{\mathcal{O}}_2)\;\widehat{\mathcal{O}}_1\big) f(x),
\end{equation}
where $f \in  L^2 (R)$. For  $\mathcal{O}_1$ and $\mathcal{O}_2$
we use the Schr\"{o}dinger representation
\begin{equation}\label{rep1}
\mathcal{O}_1 \longrightarrow \widehat{\mathcal{O}}_1 f(x):=
-i\,\hbar\,\partial_x f(x),\;\;\;\;\; \mathcal{O}_2
\longrightarrow \widehat{\mathcal{O}}_2 f(x):= \widehat{x} f(x) :=
x f(x).
\end{equation}
Thus, an explicit form of $\hat{w}$ is
\begin{equation}\label{repp1}
 \hat{w}= i\,\frac{\kappa\gamma\lambda\hbar}{2}\big(
    2 \cosh3\kappa(\phi-x)\;\frac{d}{dx}
     -3\kappa\sinh3\kappa
    (\phi-x)\big).
\end{equation}

An explicit form of $\hat{w}$ is
\begin{equation}\label{repp1}
 \hat{w}= i\,\frac{\kappa\gamma\lambda\hbar}{2}\big(
    2 \cosh3\kappa(\phi-x)\;\frac{d}{dx}
     -3\kappa\sinh3\kappa
    (\phi-x)\big).
\end{equation}
Solution to the eigenvalue problem
\begin{equation}\label{eq4}
\hat{w}\, f_a (x) = a\,f_a (x),~~~a \in R ,
\end{equation}
is found to be \cite{Malkiewicz:2009xz}
\begin{equation}\label{eq5}
f_a (x):= \frac{\sqrt{\frac{3\kappa}{\pi}}\exp\big(i \frac{2
a}{3\kappa^2 \gamma\lambda\hbar}\arctan
    e^{3\kappa(\phi-x)}\big)}{\cosh^{\frac{1}{2}}3\kappa(\phi-x)},\;\;\;\;\;\;\;\;
a =   b + 8\pi G\gamma\lambda\hbar\, m,
\end{equation}
where $b \in R$ and $m\in Z$. Completion of the span of
\begin{equation}\label{set1}
\mathcal{F}_b:=\{~f_a\;|\; a = b + 8\pi G\gamma\lambda\hbar\,
m\}\subset L^2(R),
\end{equation}
in the norm of $L^2(R)$ leads to $\infty$-dim   separable Hilbert
space $\mathcal{H}_b$. One may show \cite{ppw2} that the operator
$\hat{w}$ is essentially self-adjoint on each orthonormal space
$\mathcal{F}_b$.

Due to the the relation (\ref{vol}) and the spectral theorem on
self-adjoint operators we get the solution of the eigenvalue of
the  volume operator
\begin{equation}\label{sp1}
 v = |w|~~~\longrightarrow~~~\hat{v} f_a :=  |a| f_a .
\end{equation}
The spectrum is  bounded from below and  discrete. There exists
the minimum gap $ \bigtriangleup := 8\pi G\gamma\hbar\,\lambda\;$
in the spectrum, which defines a  quantum of the volume. In the
limit $ \lambda \rightarrow 0$, corresponding to the  classical
FRW model, there is no quantum of the volume.

The discreteness  may translate at the semi-classical level into a
foamy structure of spacetime. Our results suggest that the foamy
structure of space is a real property of the Universe so its
identification via astro-cosmo observations has sound motivation
and is  important for the fundamental physics.

\section{ Conclusions}

Modification of gravitational part of classical Hamiltonian,
realized by making use of the loop geometry (parameterized by $
\lambda$), turns big bang into big bounce (BB). Since there is no
specific $0< \lambda \in R$, there is  no specific energy density
at BB, and no specific quantum of the volume.  The spectra of
operators are parametrized by a free parameter $\lambda$ that has
not been determined theoretically, but is expected to be fixed by
the data of observational cosmology.

\begin{acknowledgement}
The author would like to thank the organizers for inspiring
atmosphere at the Meeting.
\end{acknowledgement}

\end{document}